# Alloying-Controlled Tuning of Interfacial Spin–Orbit Interaction and Magnetic Damping in Crystalline FeCo Alloys

Hongrui Lao[1], Matthias Kronseder[2], Zhe Yuan[3], Thomas Narr[1], Thomas N. G. Meier[1,5], Nadine Mundigl[2], Christian H. Back[1,4,5], and Lin Chen[1,5]

[1]*Department of Physics, TUM School of Natural Sciences, Technical University of Munich, Munich, Germany*

[2]*Institute of Experimental and Applied Physics, University of Regensburg, 93049 Regensburg, Germany*

[3]*State Key Laboratory of Surface Physics and Interdisciplinary Center for Theoretical Physics and Information Sciences, Fudan University, Shanghai 200433, China*

[4]*Munich Center for Quantum Science and Technology (MCQST), Munich, Germany*

[5]*Center for Quantum Engineering (ZQE), Technical University of Munich, Munich, Germany*

**The discovery of intrinsic spin–orbit fields in non-centrosymmetric ferromagnets has attracted considerable interest for both fundamental studies and technological applications. However, once such materials are synthesized, the strength of the spin-orbit fields is difficult to tune because it is primarily a bulk property. Here, we demonstrate that the interfacial spin-orbit interaction (SOI) in single-crystalline $Fe_{1-x}Co_x$ thin films grown on GaAs(001) can be continuously tuned via alloying. Using spin-orbit ferromagnetic resonance, we find that the Landé $g$-factor, the Gilbert damping $\alpha$, and the interfacial spin-orbit fields exhibit a common nonmonotonic dependence on Co concentration. A pronounced minimum**

**occurs near $x \approx 0.2$ where an ultra-low damping $\alpha \approx 0.0015$ is achieved. Furthermore, we observe linear scaling between $\alpha$ and $(g - 2)^2$, establishing a direct correlation between interfacial SOI and magnetic relaxation. These results identify alloying as an effective knob to engineer interfacial SOI and damping in single crystalline ferromagnet/semiconductor heterostructures.**

## 1. Introduction

Spin-orbit interaction (SOI) plays a central role in condensed-matter physics by coupling electronic spin and orbital degrees of freedom[1], thereby enabling spintronic functionalities such as spin-field transistors[2] and spin-orbit torques (SOT)[3]. In non-magnetic crystalline solids with broken inversion symmetry, SOI reshapes electronic band structures e.g. by lifting spin degeneracies and generating nontrivial spin textures in momentum space[4]. In ferromagnetic metals, SOI governs a wide range of static and dynamic magnetic properties, including the magneto-crystalline anisotropy[5], the gyromagnetic ratio[6-8], the magnetic damping[9], as well as magneto-transport phenomena such as the anisotropic magneto-resistance effect[10-12] and the anomalous Hall effect[13,14]. In single crystalline ferromagnetic metals with broken inversion symmetry either in the bulk[15-19] or at interfaces[20], SOI gives rise to the intrinsic field-like and damping-like SOTs via the inverse spin galvanic effect (iSGE)[3]. In this process, the torques are generated internally and act directly on the ferromagnetic layer. These self-induced SOTs make such systems particularly attractive, as no additional conducting layer is required and the contributions from the spin Hall effect can be excluded. iSGE induced SOTs have been demonstrated in the ferromagnetic semiconductor GaMnAs[15-18], in the Heusler alloy NiMnSb[19] and in the single crystalline ferromagnet/semiconductor hybrid Fe/GaAs (001)[20]. For the non-centrosymmetric materials GaMnAs and NiMnSb, the SOTs predominantly originate from bulk inversion asymmetry, whereas in Fe/GaAs (001) the SOTs are of interfacial origin[21-23]. Despite the observation of robust SOTs in these systems, a systematic and materials-intrinsic approach to tune the magnitude of SOTs, and thus the strength of SOI, remain largely unexplored[24]. Moreover, the influence of the SOI on magnetic relaxation is still not fully understood.

The well-known Slater-Pauling curve[25,26] describes how alloying a 3d ferromagnetic transition metal with another element shifts the Fermi energy and modifies the occupation of the magnetic $d$ states, and it has been widely used to manipulate the magnitude of the magnetic moment of ferromagnetic metals. In this article, we report that changing the $Fe_{1-x}Co_x$ composition can effectively modulate the interfacial SOI at the single crystalline $Fe_{1-x}Co_x$/GaAs heterostructure. Here, the interfacial SOI refers to the Bychkov-Rashba- and Dresselhaus-like SOIs induced by the reduced $C_{2v}$ symmetry at the interface[23]. We show that the SOI-related quantities, including the interfacial spin-orbit magnetic-fields, the Landé g-factor and the Gilbert damping, are strongly correlated and exhibit a similar nonmonotonic dependence on the Co concentration, with a local minimum at approximately 20% Co. In addition, we establish a direct link between interfacial SOI and magnetic relaxation.

## 2. Samples and spin-orbit ferromagnetic resonance

A series of 5 nm thick single-crystalline $Fe_{1-x}Co_x$ thin films with varying Co concentrations $x$ is grown by molecular-beam epitaxy (MBE) on GaAs(001) substrates. First, a 100 nm thick undoped GaAs buffer layer is deposited onto a semi-insulating

GaAs(001) substrate at a substrate temperature $T_s$ = 560 ºC. A clear (2×4) surface-reconstruction reflection high-energy electron diffraction (RHEED) pattern is observed, indicating an As-terminated flat surface. Subsequently, the GaAs substrate is transferred *in-situ* to a second MBE chamber, where 5 nm thick $Fe_{1-x}Co_x$ layers are deposited at $T_s$ = 30 ºC. Streaky RHEED patterns show a good epitaxial relationship between $Fe_{1-x}Co_x$ and GaAs for all $x$ with the films crystallizing into the same bcc phase as pure Fe. To ensure consistent growth conditions, the deposition rate of the iron cell is fixed to 0.5 nm/min, while the Co flux is adjusted to achieve the desired alloy composition. The Co concentration is limited to below 50% to stay within the bcc phase of the $Fe_{1-x}Co_x$ thin films[27,28]. Finally, a 3-nm thick Al capping layer is deposited to prevent oxidation of the $Fe_{1-x}Co_x$ films. After removal from the MBE chamber, the concentration of Co is independently verified by X-ray photoelectron spectroscopy.

We quantify the magnetic properties and the interfacial SOI using spin-orbit ferromagnetic resonance (SOFMR)[20] measurements performed on two-terminal micro-bars (20 μm × 100 μm) patterned along the [100]-direction of GaAs. An alternating microwave current flowing in the ferromagnetic layer generates a time-dependent non-equilibrium spin accumulation via the iSGE. Note that, because the GaAs substrate is electrically insulating, bulk spin Hall contributions[29] are eliminated, making $Fe_{1-x}Co_x$/GaAs a model system for isolating interfacial SOI effects. The time-dependent spin accumulation can be viewed as oscillating interfacial effective spin-orbit magnetic fields **h** (SOFs), which drive magnetization dynamics. The resulting magnetization precession produces a time-dependent resistance via the anisotropic magnetoresistance effect, leading to a rectified dc voltage at ferromagnetic resonance. This detection scheme allows simultaneous and quantitative extraction of magnetic anisotropies, the Landé g-factor, Gilbert damping, and interfacial SOFs within a single experimental platform (Fig. S1, Supporting Information).

## 3. Co concentration dependence of interfacial SOFs

Because the directions of the spin accumulation induced by Bychkov-Rashba and Dresselhaus SOIs are orthogonal in devices patterned along the [100]-direction[20,21,23], the total in-plane $h^I$ and out-of-plane $h^O$ SOFs can be expressed as $h^I=\sqrt{(h_D^I)^2+(h_R^I)^2}$ and $h^O=\sqrt{(h_D^O)^2+(h_R^O)^2}$, where $h_D^I$ ($h_R^I$) denotes the in-plane SOF arising from Dresselhaus (Bychkov-Rashba) SOI, and $h_D^O$ ($h_R^O$) the corresponding out-of-plane component (Fig. S2, Supporting Information). Microscopically, $h^I$ originates from the in-plane spin accumulation at the Fermi level $E_F$, whereas $h^O$ is generated by the combined effect of exchange coupling and spin accumulation below $E_F$, i.e., contributions from the entire occupied bands.[30] In the language of SOTs, $h^I$ corresponds to the field-like torque and modifies the magnetization precession frequency; while $h^O$ corresponds to the damping-like torque which is central to magnetization switching as well as the modulation of damping. Figures 1A and 1B show the current $I$ dependence of $h^O$ and $h^I$, respectively, measured for $x$ = 0%, 20% and 47%. Both $h^O$ and $h^I$ scale linearly with $I$, as expected from

$$h^{I(O)}=c^{I(O)}I. \qquad (1)$$

Here, the coefficients $c^O$ and $c^I$ quantify the charge-to-spin conversion efficiency and are both proportional to the strength of the effective SOI $A$, i.e., $A=\sqrt{\alpha^2+\beta^2}$, where $\alpha$ and $\beta$ characterize the strengths of the Bychkov-Rashba and Dresselhaus SOI at the interface. The Co concentration dependence $c^O$ and $c^I$ is summarized in Fig. 1C. For all Co concentrations, the magnitude of $c^I$ is approximately 3-5 times smaller than $c^O$, being consistent with previous results for Fe/GaAs and confirming the distinct microscopic origins of $h^O$ and $h^I$.[20] Moreover, both $c^O$ and $c^I$ exhibit a pronounced nonmonotonic dependence on $x$: they are largest for pure Fe, exhibit a minimum near $x \sim 20\%$, and increase again with further Co content. These results fully demonstrate efficient tuning of the interfacial SOI through alloying.

## 4. Co concentration dependence of Gilbert damping and Landé g-factor

Having analyzed the Co-concentration dependence of the current-induced spin-orbit fields, we now turn to magnetic damping and the Landé $g$-factor; both are related to SOI. Figures 2A-2C show the dependence of the FMR linewidth $\Delta H$ on the excitation frequency $f$ for $x$ = 5.9%, 20%, and 47%, respectively. For each composition, the magnetization **M** is aligned along the $<110>-$ and $<\bar{1}10>-$orientations. In all cases, a linear dependence of $\Delta H$ on $f$ is observed. The magnitude of the Gilbert damping $\alpha$ can be determined from

$$\mu_0 \Delta H = 2\alpha (2\pi f / \gamma) + \mu_0 \Delta H_0, \tag{2}$$

where $\Delta H_0$ represents the inhomogeneous linewidth broadening. For $x$ = 5.9% (Fig. 2A), the slopes of $\Delta H(f)$ along $<110>$ and $<\bar{1}10>$ directions are nearly identical, indicative of an isotropic damping, i.e., $\alpha_{<110>} \sim \alpha_{<\bar{1}10>}$. As $x$ increases to 20% (Fig. 2B), a larger slope - and thus a higher damping - is observed along $<110>$ direction, indicating the emergence of anisotropic damping with $\alpha_{<110>} > \alpha_{<\bar{1}10>}$. Interestingly, as $x$ further increases to 47% (Fig. 2C), the damping value becomes isotropic again. It is worth mentioning that the symmetry of this anisotropic damping is the same as that observed in ultra-thin Fe films grown on GaAs.[31,32]

The extracted damping parameters, $\alpha_{<110>}$ and $\alpha_{<\bar{1}10>}$, for all samples are summarized in Fig. 3A and it can be observed that the dependence of damping on $x$ also exhibits a nonmonotonic behavior. Starting from pure Fe, $\alpha$ decreases with increasing $x$ and reaches a minimum at approximately 15% Co. As $x$ further increases, $\alpha$ increases again. Notably, the lowest damping value obtained is $\alpha \sim 0.0015$, which is as low as the ultra-low damping values previously reported in other ferromagnetic metals.[33,34] The observation of such a low damping indicates that the extrinsic effects, e.g., two-magnon scattering and mosaicity broadening, do not play a significant role in the present samples. The damping anisotropy, $\Delta\alpha = (\alpha_{<110>} - \alpha_{<\bar{1}10>})/\alpha_{<\bar{1}10>}$, is presented in the inset of Fig. 3A. $\Delta\alpha$ increases with Co concentration and reaches a maximum value of 17% for $x = 10\%$.

According to Kamberský's theory of magnetic damping[9], $\alpha$ is proportional to the density of states at the Fermi level $N(E_F)$ and to the square of the spin-orbit interaction $A$, i.e.,

$$\alpha \sim N(E_F)A^2. \qquad (3)$$

Since the Gilbert damping and the interfacial SOI show a similar tendency with Co concentration, we infer that the interfacial SOI also plays an important role for damping. In fact, the Co concentration dependence of damping in polycrystalline $Fe_{1-x}Co_x$ films has previously been investigated by broadband FMR[35] and time-resolved magneto-optic Kerr effect[36]. In both studies, a local minimum in the damping has been observed for a Co concentration of approximately 20%. In the work by Schoen *et al.*, the Co-concentration dependence of damping is explained in terms of $N(E_F)$, which also shows a minimum near 20% Co[35,37]. In contrast, Mohan *et al.* demonstrates that electron-phonon scattering provides an additional contribution to damping[36,38]. In the following, we show that, in addition to the effects of $N(E_F)$ and electron-phonon scattering, the interfacial SOI indeed plays a significant role in determining the magnetic damping in $Fe_{1-x}Co_x$/GaAs (001).

In ferromagnetic metals, the spin-orbit interaction $A$ modifies the relative population of the *d*-bands sublevels and gives rise to a finite orbital moment $\mu_L$, which is proportional to $A$. The Landé *g*-factor is related to the ratio of $\mu_L$ and the spin moment $\mu_S$ via[6-8]

$$g = 2 + 2\frac{\mu_L}{\mu_S}. \tag{4}$$

Figure 3B summarizes the Co-concentration dependence of the *g*-factor obtained for $Fe_{1-x}Co_x$/GaAs (001) samples. The *g*-factor exhibits a nonmonotonic dependence on Co concentration: starting from $g \sim 2.08$ for pure Fe, it increases to a maximum value of $g \sim 2.15$ at $x = 5\%$, then it decreases to a minimum of $g \sim 2.05$ near $x = 20\%$, and then it rises again at higher Co concentrations. Since the magnitude of $\mu_L$ and $\mu_s$ in bulk $Fe_{1-x}Co_x$ is quantitatively well established[39], we extract these values to calculate the corresponding *g* values via Equation (4). As shown in Fig. 3B, the calculated *g* increases monotonically with increasing *x*, which is in clear contrast to the experimental observation. This discrepancy indicates that the Landé *g*-factor in $Fe_{1-x}Co_x$/GaAs samples is not determined by the bulk SOI of $Fe_{1-x}Co_x$ or at least not by the bulk SOI of $Fe_{1-x}Co_x$ alone.

On the other hand, Equations (3) and (4) imply a general linear relationship between $\alpha$ and $(g-2)^2$. To examine this relationship, we plot $\alpha$ as a function of $(g-2)^2$ for all the $Fe_{1-x}Co_x$/GaAs samples in Fig. 3C, where a clear linear dependence is observed. This result provide compelling evidence that interfacial SOI at the $Fe_{1-x}Co_x$/GaAs interface is a significant source of the magnetic damping, beyond the conventional contributions from the electronic density of states and electron–phonon scattering. To further validate this interpretation, we include data from our previously studied ultrathin Fe films grown on GaAs[31], in which both $g$ and $\alpha$ increase as Fe thickness is reduced. This behavior reflects the fact that Fe atoms, on average, experience the enhanced interfacial spin-orbit interaction with decreasing film thickness. As also plotted in Fig. 3C, $\alpha$ again scales linearly with $(g-2)^2$, providing additional evidence that the interfacial SOI plays an important role in determining the Gilbert damping for both $Fe_{1-x}Co_x$/GaAs and Fe/GaAs.

## 5. Discussion and conclusion

We now discuss the possible mechanism underlying the modulation of the spin–orbit fields. In general, the strength of spin–orbit interaction increases with atomic number. Although Co has a larger atomic number than Fe, the SOI strength in $Fe_{1-x}Co_x$ alloys does not vary monotonically with Co concentration. The observed nonmonotonic behavior of both the Gilbert damping $\alpha$ and the Landé $g$-factor suggests that the electronic states at the Fermi level play a decisive role in determining the effective SOI. This interpretation is consistent with the well-known contrast between Pt and Au. Despite its larger atomic number, Au exhibits relatively weak SOI-mediated transport effects because the states at the Fermi level are dominated by $s$-band electrons, resulting in a small spin Hall angle and a long spin-diffusion length.[40] In contrast, the $d$-electron character at the Fermi level of Pt leads to strong SOI effects. By analogy, we attribute the nonmonotonic SOI strength in $Fe_{1-x}Co_x$ to composition-dependent hybridization of conduction-electron states near the Fermi energy.

We further note that the Co-concentration dependence of the damping anisotropy in $Fe_{1-x}Co_x$/GaAs (001) differs significantly from previous results for $Co_{50}Fe_{50}$/MgO (001), where a giant Gilbert damping anisotropy of approximately 400% - defined as the difference between the <100> and <110> orientations -was reported and attributed to anisotropic SOI[41]. In contrast, our $Co_{50}Fe_{50}$/GaAs sample exhibits nearly isotropic damping, with the maximum anisotropy occurring at 10% Co. The alloy-induced anisotropic damping may arise from magnetization-direction-dependent modifications of the electronic density of states at the Fermi level and/or of the effective SOI. In addition, the substrate may also introduce some extrinsic influence.[42] A comprehensive understanding of the Co-concentration dependence of SOI and damping anisotropy will require detailed theoretical investigations.

In summary, we have systematically quantified the Landé $g$-factor, the Gilbert damping, and the interfacial spin–orbit interaction in binary $Fe_{1-x}Co_x$ thin films grown on GaAs(001). All three quantities exhibit a similar nonmonotonic dependence on Co concentration, with a pronounced minimum near 20% Co. These results establish the variation of the Co content as an effective tuning parameter for controlling magnetic anisotropies, the gyromagnetic ratio, anisotropic damping, and interfacial spin-orbit interaction in this material system. Since these parameters are closely linked to spin-dependent functionalities, our findings suggest that Fe$_{1-x}$Co$_x$/GaAs heterostructures provide a versatile platform for spintronic applications, including spin-orbit-torque-driven modulation of magnetization dynamics[43], control of (anisotropic) Gilbert damping, and current-induced magnetization switching.

## 6. Experimental section

*Device fabrication*: 20 μm × 100.0 μm stripes with the long side along [100]-orientation are defined by mask-free writer and ion beam etching. The contacts are made from 15 nm Ti and 150 nm Au. The stripes have a resistance ~ 600 Ω. During the fabrication, the highest baking temperature is 110 ºC.

*SOFMR measuements*: Microwave currents with a frequency of 12 GHz and the input microwave power of 22 dBm (~158 mW) are used for the angular measurements. A Bias Tee is used to separate the dc voltage from microwave background. Since the skin depth at 12 GHz is much larger than the Fe$_{1-x}$Co$_x$ thickness, the current density in Fe$_{1-x}$Co$_x$ is expected to be spatially uniform. Thus, the Oersted field should not produce net torque on Fe$_{1-x}$Co$_x$ itself. All measurements are performed at room temperature. The analysis method of the SOFMR signal is presented in Supporting Information.

## References

1. I. Žutić, J. Fabian, S. Das Sarma, *Rev. Mod. Phys.* **2004**, *76*, 323

2. F. Eberle, *et al*. *Nano. Lett*. **2023**, *23*, 4815-482.

3. A. Manchon, *et al*. *Rev. Mod. Phys*. **2019**, *91*, 035004.

4. Y. A. Bychkov, & E. I. Rashba, J. Phys. C: *Solid State Phys*. **1984**, *17*, 6039-6045.

5. S. Chikazumi, Physics of Ferromagnetism (Oxford University, Oxford, 1997), 2$^{nd}$ ed.

6. C. Kittel, Phys. Rev. **1949**, *76*, 743.

7. J. M. Shaw, R. Knut, A. Armstrong, S. Bhandary, Y. Kvashnin, D. Thonig, E. K. Delczeg-Czirjak, O. Karis, T. J. Silva, E. Weschke, H. T. Nembach, O. Eriksson, and D. A. Arena, Phys. Rev. Lett. **2021**, *127*, 207201.

8. J. Shao, *et al*. Nano. Lett. 2026, *Nano Lett*. **2026**, 26, 2853–2860.

9. V. Kamberský, Czech. J. Phys. B **1976**, *26*, 1366-1383.

10. I. A. Campbell, A. Fert, and O. Jaoul, J. Phys. C: *Solid State Phys*. **1970**, *3*, S95.

11. T. R. Mcguire and R. I. Potter, IEEE Trans. Magn. **1975**, *11*, 1018-1037.

12. F. L. Zeng, *et al*., Phys. Rev. Lett. **2020**, *125*, 097201.

13. N. Nagaosa, *Rev. Mod. Phys*. **2010**, *82*, 1539.

14. J. Guo, *et al*., arXiv **2026**, 2601.03744v1.

15. A. Chernyshov, *et al*. *Nature Phys*. **2009**, *5*, 656-659.

16. M. Endo, *et al*. *Appl. Phys. Lett*. **2010**, *97*, 222501.

17. D. Fang, *et al*. *Nature Nanotech*. **2011**, *6*, 413-417.

18. H. Kurebayashi, *et al*. *Nature Nanotech*. **2014**, *9*, 211-217.

19. C. Ciccarelli, *et al*. *Nature Phys*. **2016**, *12*, 855-860.

20. L. Chen, *et al*. *Nature Commun*. **2016**, *7*, 13802.

21. Fabian, J. *et al*. Semiconductor spintronics. *Acta Physica Slovaca*. **2007**, *57*, 565-907.

22. J. Moser, *et al*. Phys. Rev. Lett. **2007**, *99*, 056601

23. M. Gmitra, A. Matos-Abiague, C. Draxl, and J. Fabian, Phys. Rev. Lett. **2013**, *111*,

036603.

24. L. Chen, M. Gmitra, M. Vogel, *et al.* *Nat Electron* **2018**, *1*, 350–355.

25. H. P. J.Wijn, ed., *Magnetic Properties of Metals: d-Elements, Alloys and Compounds* (Springer, Berlin, 1991)

26. K. Schwarz, *et al*. J. Phys. F: Met. Phys. **1984**, *14*, 2659-2671.

27. M. A. W. Schoen, *et al*. Phys. Rev. B **2017**, *95*, 134411.

28. M. A. W. Schoen, *et al*. Phys. Rev. B **2017**, *95*, 134410.

29. J. Sinova, *et al*. *Rev. Mod. Phys.* **2015**, *87*, 1213.

30. A. Qaiumzadeh, *et al*. *Phys. Rev.* B **2015**, *92,* 014402.

31. L. Chen, *et al*. Phys. Rev. Lett. **2023**, *130*, 046704.

32. L. Chen, S. Mankovsky, S. Wimmer, M. A. W. Schoen, H. S. Körner, M. Kronseder, D. Schuh, D. Bougeard, H. Ebert, D. Weiss, and C. H. Back, Nature Phys. **2018**, *14*, 490-494.

33. Y. Wei, *et al*. Sci. Adv. **2021**, *7*, eabc5053.

34. C. Guillemard, *et al*. Phys. Rev. Appl. **2019**, *11*, 064009.

35. M. Schoen, D. Thonig, M. Schneider, *et al.* Nature Phys. **2016**, *12*, 839–842.

36. R. Mohan, *et al*. arXiv. **2021**, 2107.11699

37. S. Mankovsky, *et al*. Phys. Rev. B **2013**, *87*, 014430.

38. K. Moseni, *et al*. Phys. Rev. B **2024**, *8*, 114425.

39. S. Chadov, *et al*. EPL **2008**, *82*, 37001.

40. R. S. Nair, *et al.* Phys. Rev. Lett. **2021**, *126*, 196601.

41. Y. Li, *et al*. Phys. Rev. Lett. **2019**, *122*, 117203.

42. B. Khodadadi, *et al*. Phys. Rev. Lett. **2020**, *124*, 157201.

43. L. Chen, *et al*. Nature **2024**, *633*, 548-553.

# Figure captions

Figure 1. (A) and (B): $I$-dependence of out-of-plane SOF $h^O$ and in-plane SOF $h^I$. (C) Co-dependence of charge-to-spin conversion efficiency $c^O$ and $c^I$.

Figure 2. $f$-dependence of $\Delta H$ for $x = 5.9\%$ (A), $x = 20\%$ (B) and $x = 47\%$ (C). Here $\Delta H$ represents full width at half maximum.

Figure 3. (A) $\alpha$ and (B) $g$-factor as a function of Co concentration for **H** // <110>- and **H** // <$\bar{1}$10>-orientations. The inset of (A) shows the damping anisotropy vs. $x$. (C) $\alpha$ is plotted against $(g-2)^2$ for $Fe_{1-x}Co_x$/GaAs (solid squares) and Fe/GaAs (open circles). For both sample series, we observe a proportionality between $\alpha$ and $(g-2)^2$.

# Supporting Information

Supporting Information is available from the Wiley Online Library.

# Acknowledgements

The authors thank S. Mankovsky for fruitful discussion. This work was funded by the Deutsche Forschungsgemeinschaft by TRR 360-492547816, by SFB1277-314695032, and by 570332449.

**Conflict of Interests:** The authors declare no conflict of interests.

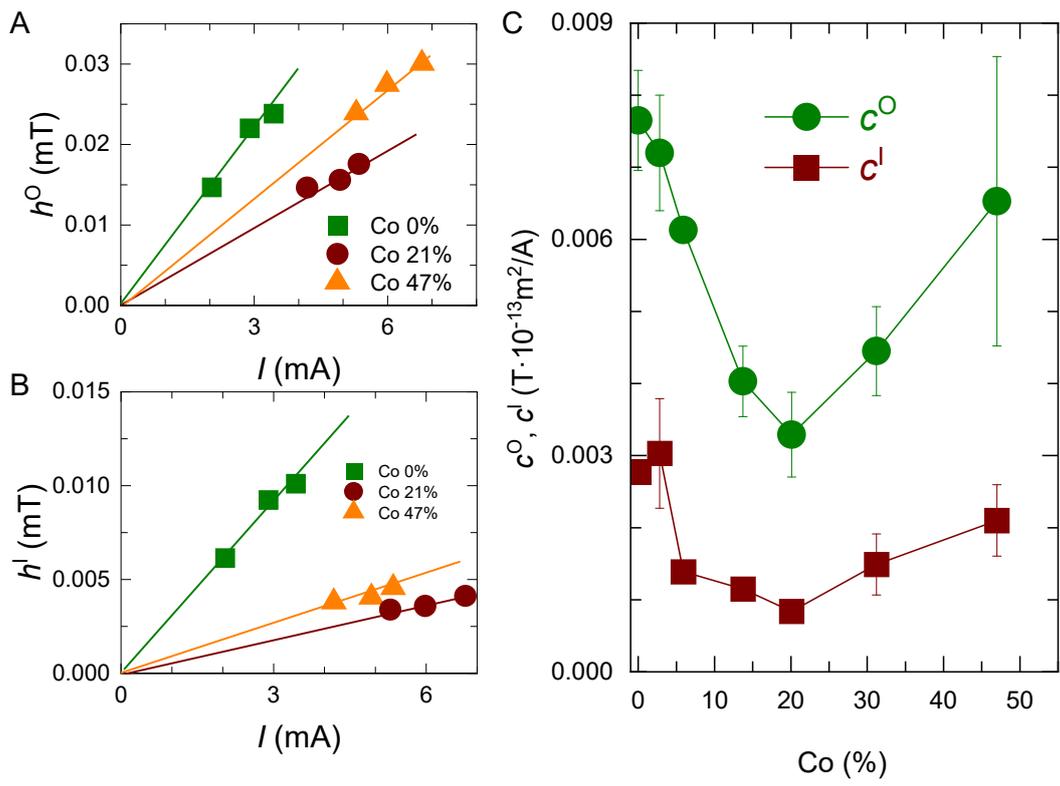

Fig. 1 Lao *et al*.

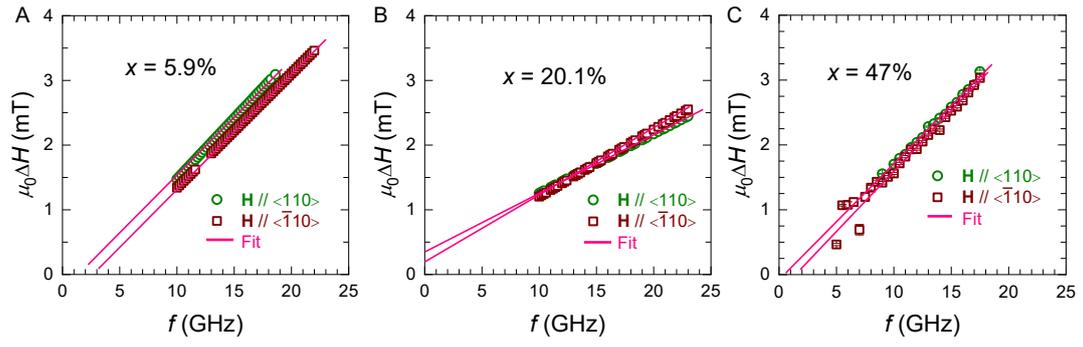

Fig. 2 Lao *et al*.

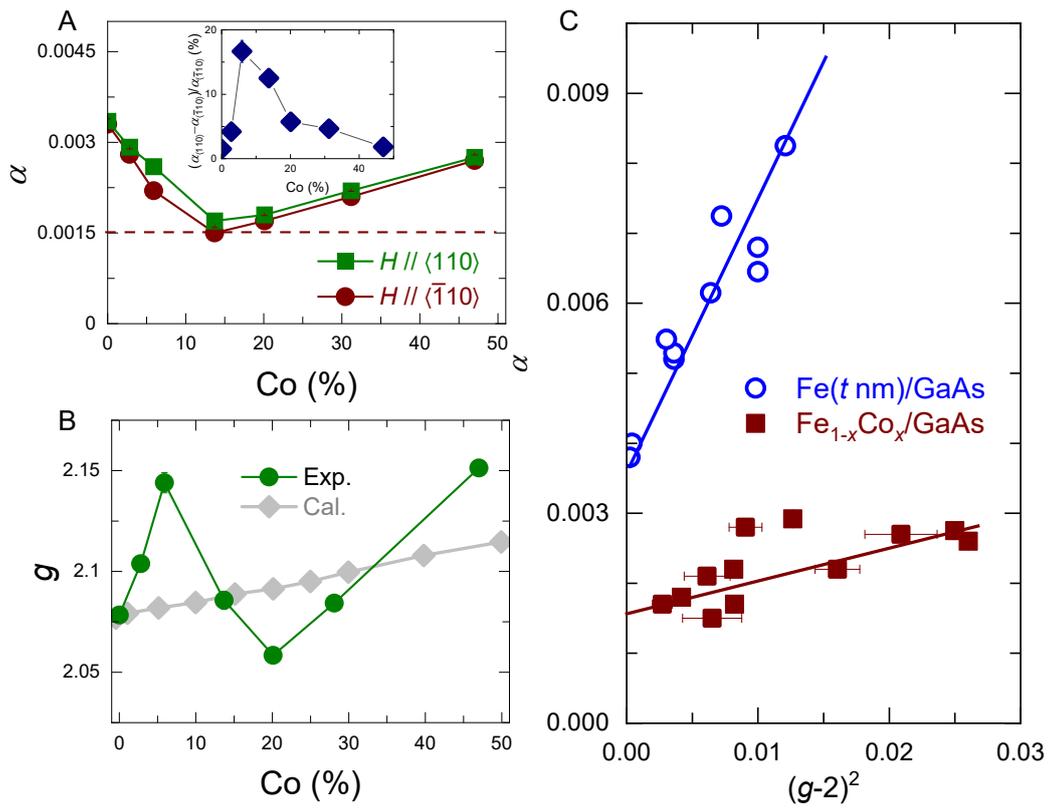

Fig. 3 Lao *et al*.